\documentclass[%
 aip,
 amsmath,amssymb,
 reprint,%
]{revtex4-1}

\usepackage{graphicx}
\usepackage{dcolumn}
\usepackage{bm}

\usepackage[utf8]{inputenc}
\usepackage[T1]{fontenc}
\usepackage{mathptmx}
\usepackage{etoolbox}
\usepackage[colorlinks=true,citecolor=blue,linkcolor=blue,urlcolor=blue]{hyperref}
\usepackage{amsmath}
\usepackage{float}

\makeatletter
\def\@email#1#2{%
 \endgroup
 \patchcmd{\titleblock@produce}
  {\frontmatter@RRAPformat}
  {\frontmatter@RRAPformat{\produce@RRAP{*#1\href{mailto:#2}{#2}}}\frontmatter@RRAPformat}
  {}{}
}%
\makeatother
\begin{document}


\title{Thermal Noise-Limited Laser Stabilization to an 8 mL Volume Fabry-P\'erot Reference Cavity with Microfabricated Mirrors}

\author{Charles~A.~McLemore}
\affiliation{Department of Physics, University of Colorado Boulder, Boulder, CO 80309, USA}
\affiliation{National Institute of Standards and Technology, Boulder, CO 80305, USA}
\email[]{charles.mclemore@colorado.edu}

\author{Naijun~Jin}
\affiliation{Department of Applied Physics, Yale University, New Haven, CT 06520, USA}

\author{Megan~L.~Kelleher}
\affiliation{Department of Physics, University of Colorado Boulder, Boulder, CO 80309, USA}
\affiliation{National Institute of Standards and Technology, Boulder, CO 80305, USA}

\author{James~P.~Hendrie}
\affiliation{Department of Physics, University of Colorado Boulder, Boulder, CO 80309, USA}
\affiliation{National Institute of Standards and Technology, Boulder, CO 80305, USA}

\author{David~Mason}
\affiliation{Department of Applied Physics, Yale University, New Haven, CT 06520, USA}

\author{Yizhi~Luo}
\affiliation{Department of Applied Physics, Yale University, New Haven, CT 06520, USA}

\author{Dahyeon~Lee}
\affiliation{Department of Physics, University of Colorado Boulder, Boulder, CO 80309, USA}
\affiliation{National Institute of Standards and Technology, Boulder, CO 80305, USA}

\author{Peter~Rakich}
\affiliation{Department of Applied Physics, Yale University, New Haven, CT 06520, USA}

\author{Scott~A.~Diddams}
\affiliation{Department of Physics, University of Colorado Boulder, Boulder, CO 80309, USA}
\affiliation{National Institute of Standards and Technology, Boulder, CO 80305, USA}
\affiliation{Department of Electrical, Computer and Energy Engineering, University of Colorado Boulder, Boulder, CO 80309, USA}

\author{Franklyn~Quinlan}
\affiliation{Department of Physics, University of Colorado Boulder, Boulder, CO 80309, USA}
\affiliation{National Institute of Standards and Technology, Boulder, CO 80305, USA}
\email[]{franklyn.quinlan@nist.gov}

\date{\today}

\begin{abstract}
Lasers stabilized to vacuum-gap Fabry-P\'erot optical reference cavities display extraordinarily low noise and high stability, with linewidths much less than 1~Hz. These lasers can expand into new applications and ubiquitous use with the development of compact, portable cavities that are manufacturable at scale. Here we demonstrate an 8~mL volume Fabry-P\'erot cavity constructed with mirrors that are fabricated lithographically with finesse near 1~million. A laser locked to the cavity exhibits phase noise limited by the cavity thermal noise for offset frequencies ranging from 1~Hz to $\approx$~1~kHz, with a fractional frequency stability of 7$\times$10$^{-15}$ at 1~second. Furthermore, the use of microfabricated mirrors allows us to expand the design space of centimeter-scale cavities, and we explore the noise implications of pushing towards cavity volumes of 2~mL or less.
\end{abstract}

\maketitle

\section{\label{sec1}Introduction}

Low-noise, highly stable Fabry-P\'erot reference cavities are integral to ultrastable laser systems used in optical clocks~\cite{ludlow2015optical, oelker2019demonstration, mcgrew2018atomic, takano2016geopotential}, optically derived low-noise microwaves~\cite{fortier2011generation, xie2017photonic}, and gravitational wave detection~\cite{willke2008stabilized}, with applications in fundamental physics~\cite{huntemann2014improved, delva2017test, derevianko2014hunting}, coherent radar~\cite{ghelfi2014fully, scheer1993}, and the redefinition of the SI second~\cite{riehle2015towards}. Through careful design with ultra-low expansion materials, Fabry-P\'erot reference cavities are engineered to maintain unparalleled length stability between their end mirrors. Locking a laser to a reference cavity transfers this length stability to the laser frequency, with state-of-the-art systems demonstrating laser linewidths below 10~mHz and fractional frequency instabilities below 10$^{-16}$~~\cite{matei20171}. This level of performance is achieved by exploiting large cavity mode volumes and/or cryogenic operation, relegating their use to well-controlled laboratory environments. However, there are a growing number of applications that demand compact, portable systems capable of low-noise operation in diverse and unpredictable environments, ranging from ground-based geodesy~\cite{grotti2018geodesy, takamoto2020test} and earthquake detection~\cite{marra2018ultrastable} to space-based tests of fundamental physics~\cite{sanjuan2019long}. For these applications, significant scientific and technical impact can be achieved with laser instability at least an order of magnitude better than what is achievable either with an unlocked laser or with a system locked to a microwave oscillator; in other words, with laser fractional instability below 10$^{-14}$ for timescales up to $\approx$~1~second. Moreover, repeatable and scalable cavity manufacture can enable widespread deployment of ultrastable laser systems, further expanding their use through integration with other chip-scale photonic devices.

\begin{figure*}
    \centering
    \includegraphics[width=\linewidth]{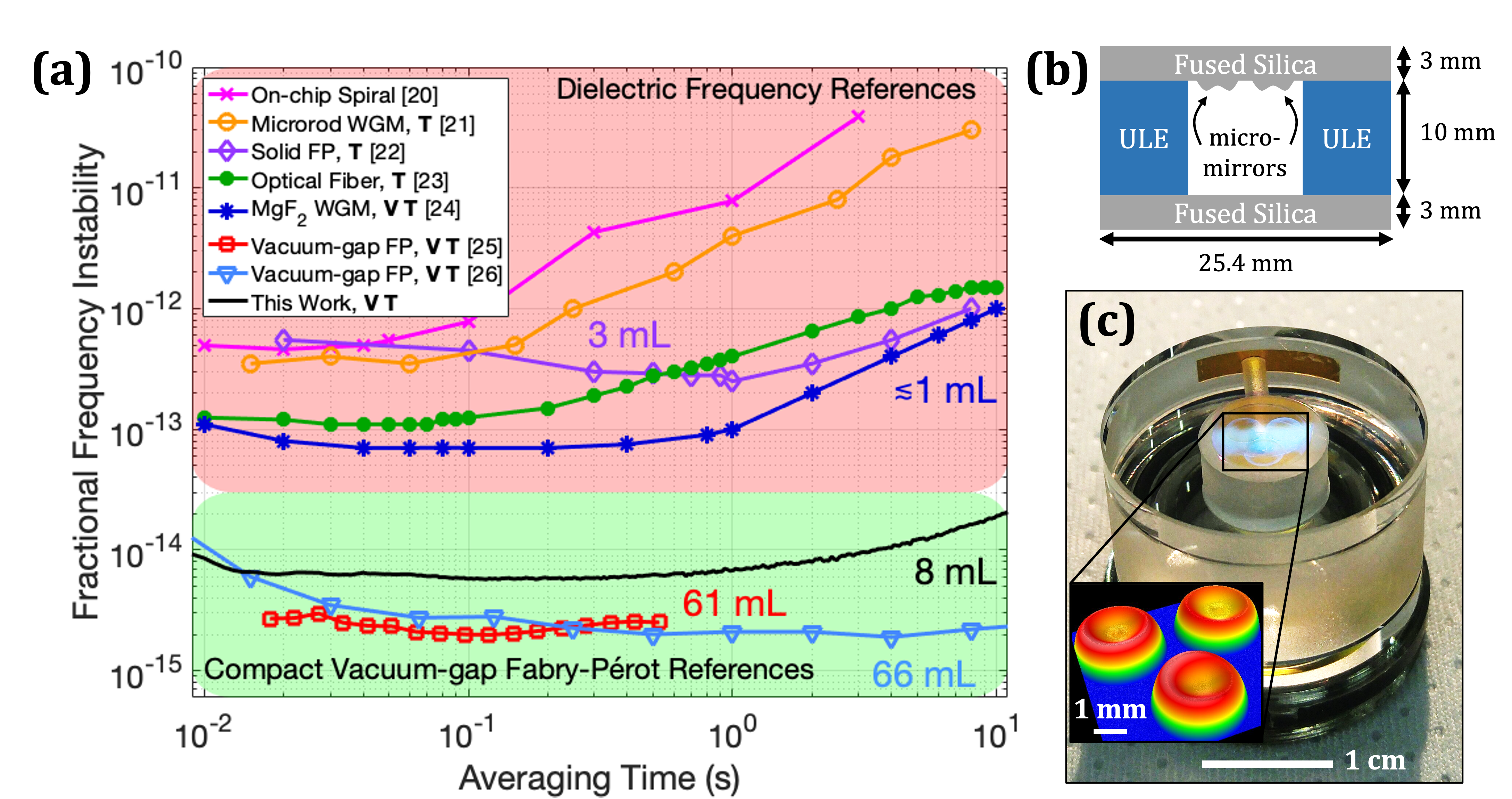}
    \caption{(a) Comparison of existing compact frequency reference technologies. FP: Fabry-P\'erot cavity, WGM: whispering-gallery-mode resonator. Bold text in legend indicates which results relied on vacuum (\textbf{V}) and/or active temperature control (\textbf{T}). (b) Cross-section of the finalized cavity geometry with an overall volume of 8~mL. (c) Photograph of the reference cavity. Inset: surface profile data of the three micromirrors fabricated on a single substrate. Color scale on inset ranges from 0~$\mu$m (blue) to 0.75~$\mu$m (red).}
    \label{fig1}
\end{figure*}

Approaches to address the need for portable ultrastable lasers can be separated into two broad categories: dielectric resonators and vacuum-gap Fabry-P\'erot cavities, each of which has its own advantages. While dielectric resonators offer compact and manufacturable device structures, their frequency stability is fundamentally limited by thermo-refractive noise~\cite{lim2017chasing}, as the optical mode is confined within material at finite temperature, and they typically have greater susceptibility to long-term thermal drift. Thus, a significant degree of active temperature control is often necessary to maintain frequency stability near the thermo-refractive noise limit. In contrast, vacuum-gap Fabry-P\'erot cavities have an optical mode that only interacts with the mirror surfaces, reducing the influence of stochastic thermal fluctuations inherent to all materials, and they can be constructed with low thermal expansion materials that greatly reduce the required temperature stability. Fig.~\ref{fig1}a shows a comparison of the fractional frequency instability of some of the most stable solid-state resonators~\cite{lee2013spiral, zhang2019microrod, stern2020ultra, loh2020operation, alnis2011thermal} relative to what has been achieved with compact vacuum-gap Fabry-P\'erot cavities~\cite{davila2017compact, leibrandt2013cavity}. The improved performance of Fabry-P\'erot cavities comes at the expense of larger cavity volume, small-batch lap and polish manufacturing, and individual hand assembly. Thus, the ideal portable ultrastable laser would inherit the frequency stability of bulk Fabry-P\'erot reference cavities but with compactness and manufacturability associated with microresonator designs. 

To this end, here we present a near 1~million finesse Fabry-P\'erot reference cavity constructed with lithographically fabricated micromirrors. The fabrication process allows for highly parallel manufacture of mirrors on a single substrate with user-defined radius of curvature. The constructed cavity, shown in schematically in Fig.~\ref{fig1}b and pictured in Fig.~\ref{fig1}c, has a volume of only 8~mL. We confirm the cavity thermal noise is not adversely affected by the manufacturing process, with the phase noise of a laser locked to the cavity operating at the predicted thermal noise limit out to $\approx$~1~kHz offset. The corresponding fractional frequency instability is close to the thermal limit of 5$\times$10$^{-15}$ and is better than 10$^{-14}$ up to several seconds of averaging. We use finite element simulation to further explore the design space, predicting thermal noise and cavity thermal expansion as a function of mirror substrate thickness. This demonstration of a high-performance optical reference cavity using microfabricated mirrors represents an essential step towards the integration of ultrastable frequency references in numerous field-based experiments.

\section{\label{sec2}Noise Modeling \& Cavity Design}

\emph{Mirror design \& fabrication.} Several compact Fabry-P\'erot cavity designs have been proposed and demonstrated to specifically address the need for portability while meeting high stability requirements~\cite{davila2017compact, leibrandt2013cavity, didier2016design, webster2011force}. These cavities exploit a high degree of mechanical symmetry to reduce acceleration and holding force sensitivity and use large radius of curvature (ROC) mirrors to reduce the cavity thermal noise limit. Importantly, these cavities use standard mirrors that are fabricated by lapping and polishing a curved surface on a glass substrate. To permit bonding of this substrate to a planar cavity spacer, it is necessary to polish a flat annulus on the outer rim of the mirror. Both of these features (i.e., curved mirror and bonding annulus) we instead create using lithographic techniques, as we describe here. 

Central to our cavity is a curved mirror formed using a novel lithographic technique that utilizes a carefully engineered photoresist reflow and etching to form mm-scale diameter micromirrors with customizable ROC~\cite{kharel2018ultra, jin2022scalable}. The process consists of patterning a disk of photoresist onto a super-polished glass substrate, which is then exposed to a solvent vapor reflow in a custom chamber. Gradually, the solvent vapor is absorbed into the photoresist, which begins to flow as surface tension reshapes the disk. The reflow process can be halted at a prescribed time by baking out the solvent from the photoresist, resulting in a small dimple shape on the top of the photoresist. This dimple approximates a parabolic surface near its center and serves as a template for a concave mirror. Fine tuning of the reflow duration along with the initial dimensions of the photoresist disk allow for a wide range of mirror curvatures to be achieved with this technique; parabolic surfaces have been produced with radii of curvature ranging from less than 1~mm to over 1~m. Following the reflow step, a reactive ion etch is used to transfer the parabolic micromirror shape into the glass substrate below with angstrom-level surface roughness. Finally, a highly reflective dielectric mirror coating is applied via ion beam sputtering to complete the micromirror. Full details of this process are available in Ref.~\onlinecite{jin2022scalable}. Since the photoresist is applied lithographically and the whole substrate is etched at once, this technique allows for a high degree of parallel fabrication. Furthermore, the bonding annulus can be lithographically defined as well, leading to a high degree of repeatability in the cavity assembly.

Since we seek to demonstrate the viability of these microfabricated mirrors for use in ultrastable frequency references, we maximized the micromirror ROC to $\approx$~1~m to correspondingly increase the optical spot size. This averages thermal fluctuations over a greater area, thereby lowering the thermal noise floor~\cite{levin1998internal}. Furthermore, to evaluate the repeatability and parallel manufacture capabilities of the micromirror fabrication technique, we included 3 individual large ROC micromirrors on a single mirror substrate, each capable of forming an optical cavity when paired with one optical flat. To optimize the remainder of the cavity design, we performed finite element simulations to assess both the thermal noise and temperature sensitivity of various geometries. With our microfabricated mirrors, the design space includes mirror substrate thicknesses much less than that available with standard polished mirrors, as well as substrates with smaller cross-sectional diameter. 

\emph{Noise modeling.} The most crucial considerations in the cavity design are those that affect the thermal noise floor, which must be below $1.3/f^3$ dBc/Hz (where $f$ is the offset frequency) to reach fractional frequency instability of below 10$^{-14}$. Thermal noise in Fabry-P\'erot reference cavities is rooted in thermally driven stochastic fluctuations in the cavity mirrors and spacer. By individually considering how thermal fluctuations affect the length of the optical axis through different mechanisms, we can calculate the thermal noise floor as the sum of these contributions. In our analysis, we consider Brownian noise arising from mechanical damping within the materials~\cite{levin1998internal, kessler2012thermal}, thermo-elastic noise resulting from the coupling of thermal fluctuations to a finite coefficient of thermal expansion (CTE) in the cavity materials~\cite{liu2000thermoelastic, cerdonio2001thermoelastic}, and thermo-refractive noise due to a temperature-dependent index of refraction within the mirror coatings~\cite{evans2008thermo}. While analytical models exist to calculate these noise sources, the models often depend on simplifying assumptions that typically hold for bulk Fabry-P\'erot cavities. However, as we design more compact cavities, some of these assumptions break down, such as assuming the mirror substrates are large compared to the characteristic length scale of thermal diffusion, and assuming the noise from the mirror substrates and spacer can be treated independently. Following Ref.~\onlinecite{kessler2012thermal}, we used finite element analysis (COMSOL) to construct a model of our cavity and apply a force to the mirrors with a Gaussian profile matching that of our optical mode. After computing the resulting deformation of the model cavity, we leverage the fluctuation-dissipation theorem~\cite{callen1951irreversibility} to extract the power spectral density of mirror surface fluctuations. Summing the contributions from Brownian, thermo-elastic, and thermo-refractive noise sources, we arrive at the total expected thermal noise of a given cavity design. As pointed out in Ref.~\onlinecite{kessler2012thermal}, the Brownian noise from a small cavity spacer can be grossly underestimated when using simplified analytical models. We additionally find that, for the thin mirror substrates available with microfabrication, the Brownian noise of the coating and mirror substrate are affected by the contact area between the mirror and spacer. A larger contact area tends to restrict the drumhead-like motion, driven by Brownian noise, that otherwise increases the mirror surface displacement. 

\emph{Cavity thermal expansion.} In addition to calculating the thermal noise floor of the cavity, we also utilized finite element simulations to determine the temperature sensitivity of various cavity designs. While stochastic thermal noise dominates the stability of a Fabry-P\'erot reference cavity at short timescales, temperature drift of the optical cavity length typically dominates over long timescales. To maintain sufficient fractional frequency stability over several seconds, it was therefore necessary to reduce the magnitude of the effective cavity CTE through careful design of the cavity geometry and materials. We focused our CTE simulations on a cavity structure that includes our test substrate with 3 micromirrors. This required a total cavity diameter of 25.4~mm, with a 10~mm diameter bore hole in the center, while the mirror substrate thickness and material were left as free parameters. We then simulated the change in length of the optical axis (sampled with a Gaussian profile matching the optical mode) as the temperature of the cavity was swept. The slope of this length change with respect to temperature is proportional to the cavity’s effective CTE~\cite{legero2010tuning}, which we used to evaluate the relative temperature sensitivity.

\begin{figure}
    \centering
    \includegraphics[width=\linewidth]{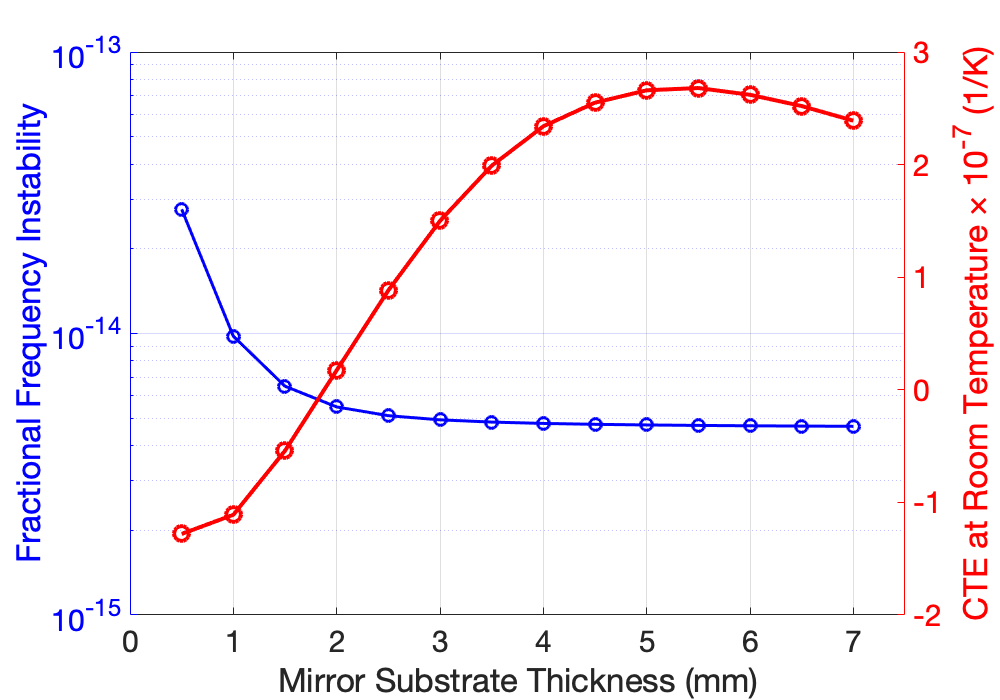}
    \caption{(a) Simulated trade-off between fractional frequency instability and cavity temperature sensitivity for various potential mirror substrate thicknesses. The CTE zero-crossing of the 10~mm ULE spacer used in the simulation is 45~$^{\circ}$C, whereas the zero-crossing of the ULE used in the experiment is specified at 45~$^{\circ}$C~$\pm$~5~$^{\circ}$C.}
    \label{fig2}
\end{figure}

Using noise and temperature sensitivity simulations in conjunction, we considered several design trade-offs to arrive at a compact, low-noise design. For example, while using fused silica as the mirror substrate material provides a lower thermal noise floor (due to the material’s high mechanical quality-factor $Q_M$ and correspondingly low internal damping), it also increases the magnitude of the effective cavity CTE when compared to other possible substrate materials like ultra-low expansion (ULE) glass. To counteract this increased cavity CTE, we explored thinner mirror substrates, which improve the CTE, although making the substrates too thin results in a significant increase to the thermal noise (Fig.~\ref{fig2}). Furthermore, we found that, for a given cavity outer diameter, increasing the bond area reduces the effective cavity CTE. It is worth noting that in future designs, the inclusion of only a single micromirror will allow for a large reduction in cavity and bore hole diameter, and therefore total volume, without sacrificing noise performance. For example, reducing the cavity diameter by half to 12.7~mm would result in a cavity volume of only 2~mL while maintaining a comparable level of thermal noise.

The ability to lithographically tailor the shape of the substrate surface allows us to overcome some of the limitations of conventional fabrication techniques. Considering the importance of the optical contact bond area on both the noise and cavity CTE, it is worth noting the level of control our microfabrication technique provides over this parameter. Two factors set the upper bound on the bond area: the spacer bore hole diameter and the inner diameter of the mirror substrate annulus. In our experience, standard curved mirrors created through lapping and polishing display significant variability in the mirror contact annulus, leading to unpredictability in CTE and thermal noise when forming bonded cavities. Additionally, large ROC mirrors have a central recess of only a few micrometers, with smaller mirror diameters yielding smaller recess for a given ROC. This puts a lower limit on the diameter of the mirror, since otherwise the annulus polishing will damage the pristine surface at the mirror’s center. In our microfabrication procedure, an annular area around the edge of the substrate is protected with a thick coating of photoresist, blocking the reactive ion etch. This leaves an intact bonding annulus of precisely controlled area. Furthermore, this annulus protection scheme guarantees that the surface quality of the bonding area remains sufficient for optical contact bonding. As we push towards building more compact cavities with thinner mirror substrates, larger ROC, and smaller overall diameter, the precise control over bond area that this technique allows will be essential for repeatably achieving target performance. Combined with the numerical simulation methods discussed above, these tools will help realize a new regime of compact cavities that push beyond traditional geometries where analytical solutions apply.

Implementing this annulus protection scheme, we finalized the cavity design with a 10 mm long ULE spacer (CTE zero-crossing: 45~$^{\circ}$C~$\pm$~5~$^{\circ}$C) to minimize thermal sensitivity while still maintaining sub-10$^{-14}$ fractional frequency instability. The minimum substrate thickness we could readily obtain with suitably low roughness (for high finesse) and high surface figure (for optical contact bonding) was 3~mm. This choice of substrate thickness also provides a good compromise between noise and CTE. The completed cavity design was assembled by optical contact bonding the fused silica mirror substrates (with dielectric coating reflectivity >99.999\%) to the ULE spacer, resulting in an overall volume of 8~mL, effective CTE of 1.5$\times$10$^{-7}$~K$^{-1}$ at room temperature, thermal phase noise floor of $-4.4/f^3$  dBc/Hz, and fractional frequency instability floor of $5~\times~10^{-15}$.

\section{\label{sec3}Experimental Results \& Discussion}

\emph{Cavity finesse.} For a compact cavity of short length, high finesse is particularly important to achieve a narrow cavity resonance, which enables greater rejection of electronic noise sources in the laser frequency lock~\cite{black2001introduction} and improves rejection of frequency instabilities due to residual amplitude modulation~\cite{zhang2014reduction}. The finesse is dependent on the mirror transmission and excess loss due to scattering, absorption, and beam clipping~\cite{hood2001characterization}. To maintain high finesse, it is critical that the micromirror etching process does not degrade the original angstrom-level roughness of the super-polished substrate, as this could lead to excess scattering losses. Furthermore, while scattering and absorption are largely intrinsic to a given mirror, determined by the roughness of the surface and quality of the dielectric coating, significant additional beam clipping loss can be introduced when rigidly bonding a cavity. Due to the mm-scale diameter of these micromirrors, even minor errors in parallelism (greater than 0.1~mrad) between the end faces of the spacer can result in significant finesse reduction due to beam clipping. Small temperature gradients in the reflow chamber while forming the micromirror shapes could also introduce asymmetries and mirror tilt, which, once bonded to a rigid spacer, would shift the position of the cavity’s optical mode transversely, incurring clipping losses as the optical mode overlaps the edge of the micromirror.

\begin{figure}
    \centering
    \includegraphics[width=\linewidth,height=0.8\textheight,keepaspectratio]{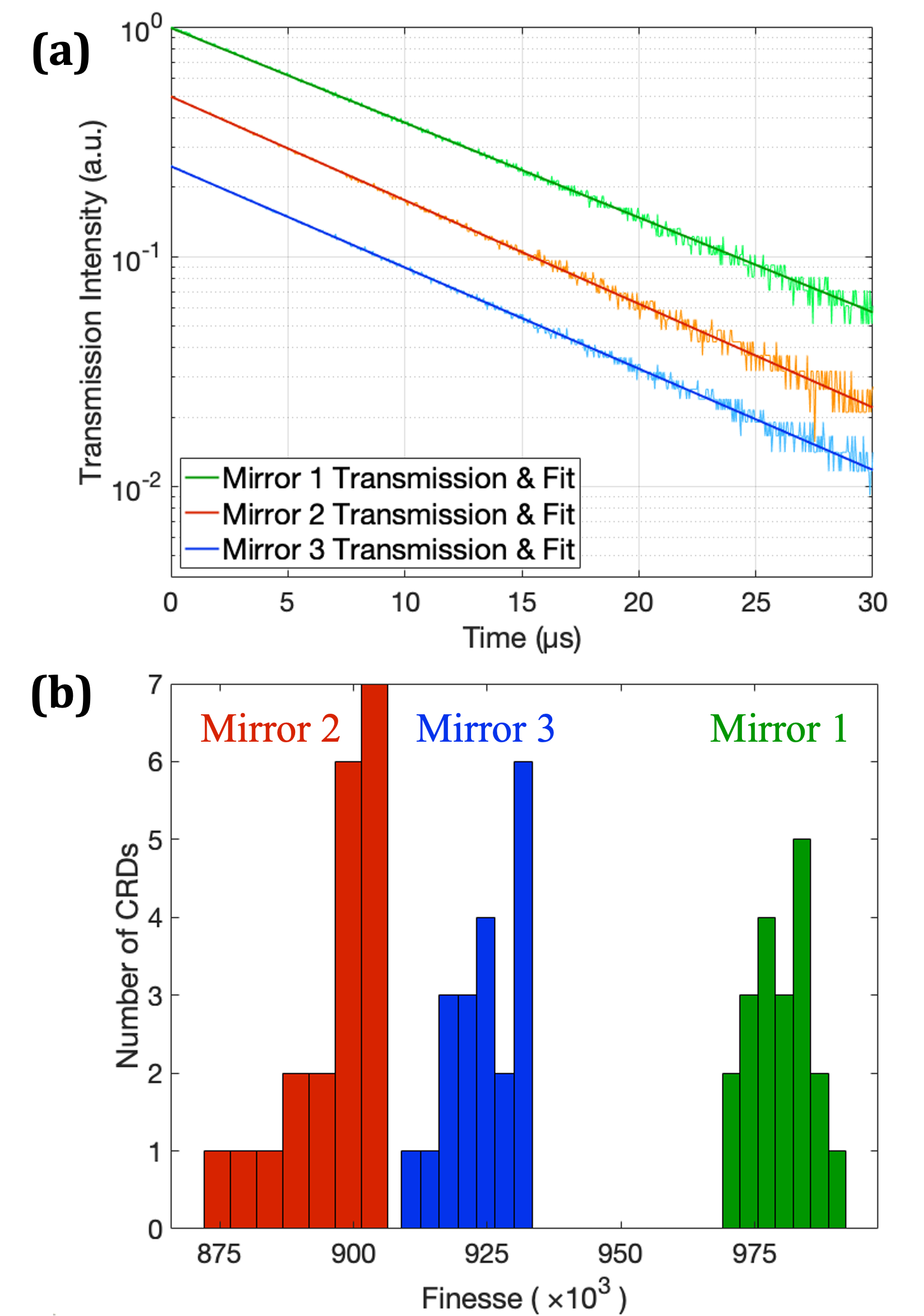}
    \caption{(a) Cavity ringdown measurements for each of the 3 micromirrors (offset for clarity), including exponential fits used to extract the photon lifetime $\tau$. (b) Histogram showing the repeatability of the CRD measurement for each micromirror, with mean values of 900~000, 920~000, and 980~000.}
    \label{fig3}
\end{figure}

In light of these considerations, we ensured that the bonded cavity maintained consistently high finesse by performing cavity ringdown (CRD) measurements on all three micromirrors~\cite{anderson1984mirror}. By measuring all three, we confirmed that the micromirrors exhibit pristine surface quality and minimal tilt asymmetries with a high degree of repeatability. To perform the CRD measurements, we swept the frequency of a commercial fiber laser through a TEM$_{00}$ mode of the cavity while sending the light through an acousto-optic modulator (AOM). When a photodetector monitoring the transmission through the cavity witnessed a spike in intensity, it triggered a high-speed switch, which cut off the radio-frequency (RF) power to the AOM, blocking the light incident on the cavity. We then monitored the exponential decay of resonant light out of the cavity using the transmission detector. With the exponential decay time constant $\tau$ (equivalent to the photon lifetime) and the cavity length $L$, we then determined the finesse as $F=\pi \tau c/L$, with $c$ the speed of light in vacuum. Fig.~\ref{fig3}a shows CRD data for all three micromirrors in this cavity, along with exponential fits, which are used to extract the photon lifetime in each case. In order to assess the repeatability of the measurement, we performed twenty CRD measurements on each micromirror. The results are presented in Fig.~\ref{fig3}b, with the average finesse of micromirrors 1, 2, and 3 being 980~000, 900~000, and 920~000, respectively. We separately measured the coating transmission to be $\approx$~1.9~ppm~\cite{jin2022scalable}, indicating single mirror loss of the contacted cavity of only $\approx$~1.3~ppm, $\approx$~1.6~ppm, and $\approx$~1.5~ppm for mirrors 1, 2, and 3, respectively. In addition to the mirrors used to form this cavity, over a dozen other mirrors were tested without bonding to spacers (allowing for optimal alignment), yielding a mean finesse value near 1~million and below 1~ppm scattering loss, with the highest finesse mirrors reaching 1.3~million~\cite{jin2022scalable}. These results confirm that the etching process does not introduce substantial scattering loss and rigid bonding of the fabricated substrates in a cavity introduces significantly less than 1~ppm of additional clipping loss per mirror. With finesse values consistently near 1~million, all three micromirrors are suitable for locking a laser at the thermal noise limit.

\emph{Phase noise \& frequency stability.} Having established the high finesse of all three micromirrors in the cavity, we chose micromirror 3 for the remaining measurements since it has the largest ROC at 1.1~m, leading to the lowest thermal noise limit, as discussed in Section~\ref{sec2}. However, this thermal noise limit in part depends on the high mechanical $Q_M$ of the underlying fused silica used as a mirror substrate. While high $Q$ phonon resonances in quartz have been demonstrated with a similar surface etch process~\cite{kharel2018ultra}, and the finesse measurements confirmed that the micromirror fabrication technique maintains the original surface quality of the substrate, it was also necessary to measure the fundamental thermal noise of the mirrors to confirm that the etch process used in the micromirror fabrication did not significantly degrade the $Q_M$ of the substrates. For our cavity, the phase noise sensitivity to changes in $Q_M$ is limited by the coating noise, estimated to be about 10~dB above the mirror substrate noise. Thus, a tenfold reduction in $Q_M$ would increase the total noise approximately 3~dB. Coatings with lower thermal noise~\cite{cole2013tenfold} or a cavity with a larger spot size would reduce the separation between coating and substrate noise and increase the sensitivity to small changes in $Q_M$ of the substrate. 

\begin{figure*}
    \centering
    \includegraphics[width=\linewidth]{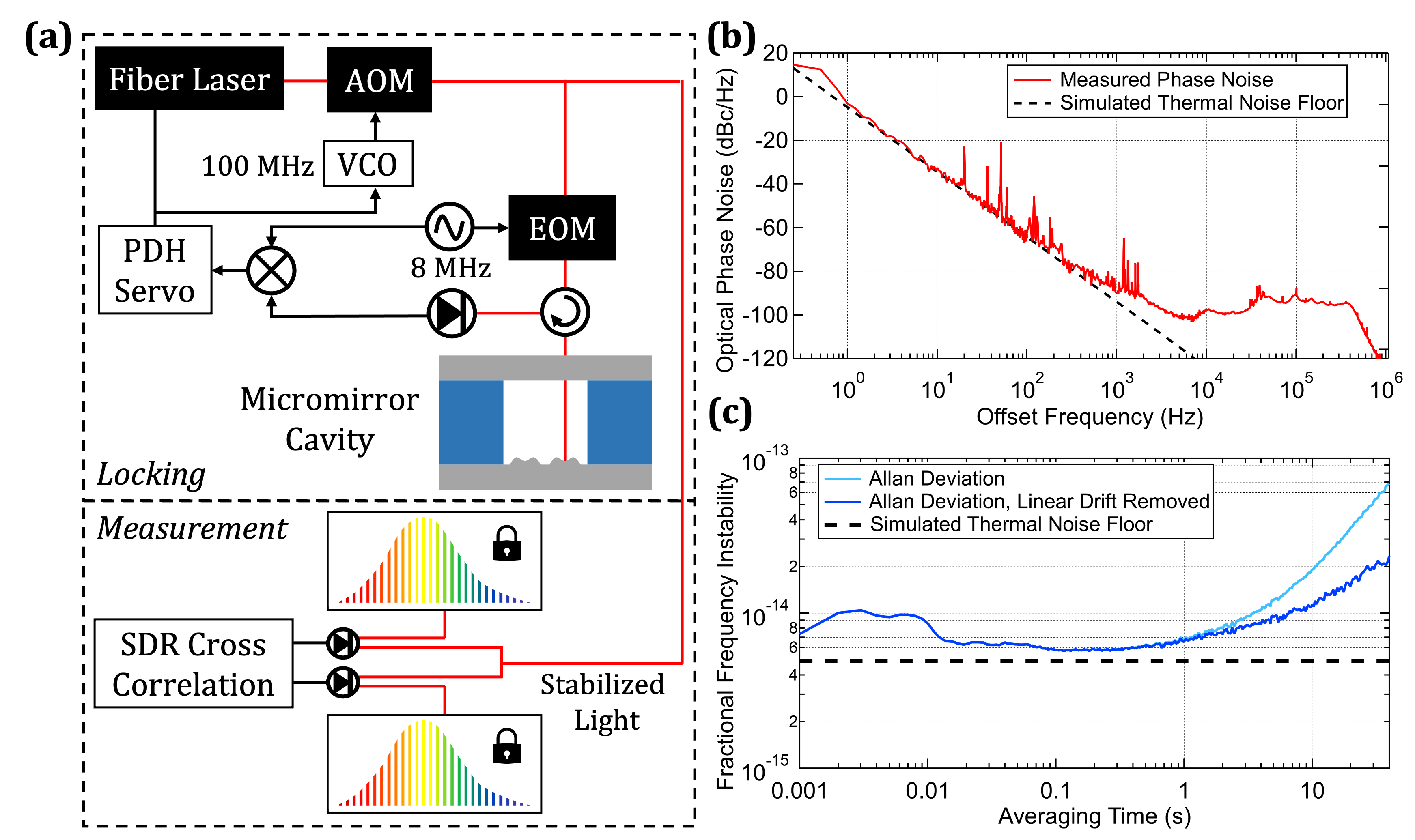}
    \caption{(a) Experimental setup for PDH locking and phase noise measurement. PDH, Pound-Drever-Hall; VCO, voltage-controlled oscillator; AOM, acousto-optic modulator; EOM, electro-optic modulator; SDR, software-defined radio. (b) Phase noise measurement of the stabilized light. Roll-off above 400~kHz is a measurement artifact. (c) Fractional frequency instability with and without removal of 1.3~Hz/s linear drift. The effective bandwidth of this measurement is 2~kHz.}
    \label{fig4}
\end{figure*}

Using the TEM$_{00}$ mode of the cavity formed by this mirror, we locked the output frequency of a fiber laser to the cavity using the Pound-Drever-Hall (PDH) method~\cite{drever1983laser}, depicted in Fig.~\ref{fig4}a. The output from a commercial fiber laser is sent through an AOM and an electro-optic modulator (EOM), which applies sidebands at 8~MHz. The light is then coupled to the cavity, which is housed in a rigid holding structure with a heat shield, held in a vacuum enclosure at 10$^{-5}$~Pa with active temperature stabilization applied to the outside of the vacuum enclosure. For further environmental isolation, the entire apparatus is contained in an acoustic damping box with active vibration cancellation. The reflected signal from the cavity is photodetected and demodulated with a mixer to extract an error signal, which is filtered and amplified to feed back to both the laser and the AOM. Fast feedback is applied to the AOM via frequency modulation implemented with a voltage-controlled oscillator (VCO), while low-pass filtered slow feedback is applied to a piezoelectric tuning port on the laser. To measure the stability of the locked laser, stabilized light is split off after the AOM and divided into two channels for comparison with two optical frequency combs stabilized to independent ultrastable optical references~\cite{nakamura2020coherent}. Heterodyne beat notes from each arm are digitally sampled with software-defined radio (SDR) at 2~MSa/s and cross-correlated to remove noise not common to both channels, leaving just the noise of the stabilized light~\cite{davila2017compact, sherman2016oscillator}. As Fig.~\ref{fig4}b shows, the phase noise of the stabilized light follows the simulated thermal noise limit of the cavity out to around 1~kHz offset, confirming that the micromirror fabrication process does not substantially alter the high mechanical $Q_M$ of the fused silica substrate and demonstrating the utility of these micromirrors for low-noise laser stabilization. The noise at higher offset frequencies is dominated by residual noise of the fiber laser. Finally, to document the longer-term stability of the cavity, the Allan deviation is extracted from the phase record of one of the two heterodyne beat notes sampled with SDR (Fig.~\ref{fig4}c). At 1~second, the Allan deviation is $7\times10^{-15}$. Following removal of a 1.3~Hz/s linear drift, the fractional frequency instability is shown to be 10$^{-14}$ or better out to 7~seconds of averaging, after which nonlinear drift dominates.

\section{\label{sec4}Conclusion}

These results demonstrate ultrastable laser performance from a compact 8~mL cavity using lithographically fabricated mirrors. Finite element noise modeling indicates that the volume can be reduced to 2~mL without sacrificing the noise performance by simply reducing the cavity’s cross-sectional diameter. With consistently high finesse across all three micromirrors, we show that this scalable microfabrication technique maintains very high surface quality, with minimal excess losses due to absorption, scattering, or clipping from poor alignment tolerances or micromirror shape. The finesse values obtained for all three micromirrors are consistent with total excess loss of less than 2~ppm per mirror. Furthermore, the thermal noise limited performance out to $\approx$~1~kHz offset confirms that the mirror fabrication technique does not significantly degrade the fused silica substrate’s naturally high mechanical $Q_M$. Finally, the fractional frequency instability reaches $6\times10^{-15}$ between 0.1~s and 1~s, and remains below 10$^{-14}$ out to several seconds of averaging, making this cavity useful for portable optical clock applications. 

While recent years have seen impressive efforts to harness some of the precision of state-of-the-art reference cavities into more portable systems, these efforts have largely relied on individually polished bulk optics. In order for a new generation of stable, portable frequency references to proliferate to a wide range of field applications, it will be necessary to make many of them, which requires a path towards scalable manufacturing. We envision the possibility of fabricating many cavities in parallel: by patterning a grid of micromirrors onto a single substrate, bonding this with a spacer possessing a matching grid of bore holes, and capping the other end with a flat mirror substrate, we could simultaneously assemble a large number of reference cavities. Then, using foundry-scale dicing techniques, the reference cavities could be divided into individual units. Here, we take a first step towards the mass manufacture of portable, ultrastable Fabry-P\'erot reference cavities by demonstrating better than 10$^{-14}$ instability and thermally limited performance over several decades using micromirrors fabricated in a highly repeatable lithographic process.

\begin{acknowledgments}
We thank Chun-Chia Chen and Amit Agrawal for helpful comments on the manuscript. We also thank Andrew Ludlow and the NIST Yb clock team for providing stable reference light. This work was funded by the DARPA A-PhI program and NIST. This work is a contribution of an agency of the U.S. government and not subject to copyright in the USA. Commercial equipment is identified for scientific clarity only and does not represent an endorsement by NIST.
\end{acknowledgments}

\section*{Data Availability Statement}

The data that support the findings of this study are available from the corresponding author upon reasonable request.


\bibliography{references}

\end{document}